\documentclass[aps,prl,preprint,superscriptaddress]{revtex4}
\usepackage{amsmath,array,graphicx}

\begin{document}
\title{Control of light transmission through opaque scattering media in space and
time}
\author{Jochen Aulbach}
\email{j.aulbach@amolf.nl} \affiliation {FOM Institute for Atomic
and Molecular Physics AMOLF, Science Park 113, 1098 XG Amsterdam,
The Netherlands} \affiliation{Institut Langevin, ESPCI ParisTech,
CNRS, 10 rue Vauquelin, 75231 Paris Cedex 05, France}
\author{Bergin Gjonaj}
\affiliation {FOM Institute for Atomic and Molecular Physics AMOLF,
Science Park 113, 1098 XG Amsterdam, The Netherlands}
\author{Patrick M. Johnson}
\affiliation {FOM Institute for Atomic and Molecular Physics AMOLF,
Science Park 113, 1098 XG Amsterdam, The Netherlands}
\author{Allard P. Mosk}
\affiliation {Complex Photonic Systems, MESA+ Institute for
Nanotechnology and Department of Science and Technology, University
of Twente, Post Office Box 217, NL-7500 AE Enschede, The
Netherlands}
\author{Ad Lagendijk}
\affiliation {FOM Institute for Atomic and Molecular Physics AMOLF,
Science Park 113, 1098 XG Amsterdam, The Netherlands}
\begin{abstract}
We report the first experimental demonstration of combined spatial
and temporal control of light trajectories through opaque media.
This control is achieved by solely manipulating spatial degrees of
freedom of the incident wavefront. As an application, we demonstrate
that the present approach is capable to form bandwidth-limited
ultrashort pulses from the otherwise randomly transmitted light with
a controllable interaction time of the pulses with the medium. Our
approach provides a new tool for fundamental studies of light
propagation in complex media and has potential for applications for
coherent control, sensing and imaging in nano- and biophotonics.
\end{abstract}

\maketitle
\begin{figure}[t!]
  \includegraphics[width=0.34\textwidth]{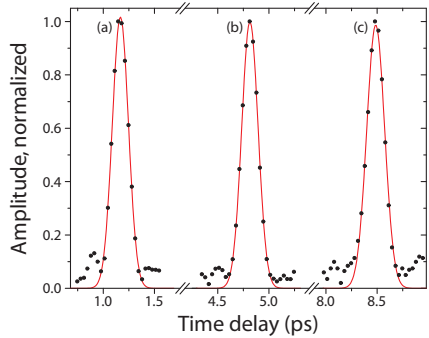}\\
  \caption{Detailed cross-correlation scans around the optimized pulses at
  the time delays $\tau_{\rm opt}$~=~1.1~ps ({a}), 4.8~ps ({b}) and
  8.5~ps ({c}), together with a Gaussian fit (red
  lines). The amplitudes have been normalized to the maximum of the respective peak.
  The width (FWHM) of the peaks are $\Delta\tau_{\rm opt}$~=~186 fs ({a}), 188 fs ({b}) and
  206 fs ({c}). These values are typical in
  the range of time delays between zero and 10~ps, with an average
  width $\Delta\tau_{\rm opt} = (190\,\pm\,7)$fs.}
  \label{FigOptimizedPeakWithFit}
\end{figure}

Concentrating light in time and space is critical for many
applications of laser light.
Broad-band mode-locked lasers provide the required ultrashort light
pulses for multiphoton imaging
\cite{hell_breaking_1994,zipfel_nonlinear_2003}, nanosurgery
\cite{vogel_mechanisms_2005}, microstructuring
\cite{glezer_three-dimensional_1996}, ultrafast spectroscopy
\cite{zewail_femtochemistry:_2000,shah_ultrafast_1999} and coherent
control of molecular dynamics or of nanooptical fields
\cite{rabitz_whitherfuture_2000,
herek_quantum_2002,aeschlimann_adaptive_2007}.
%
Multiple random scattering in complex media severely limits the
performance of these methods, but often is an unavoidable nuisance
in many systems of interest, such as biological tissue or
nanophotonic structures \cite{koenderink_light_2003}. Spatially,
random scattering strongly distorts a propagating wave front,
creating the well-known speckle interference pattern
\cite{dainty_laser_1984}. In the time domain, ultrashort pulses are
strongly distorted and widely stretched due to the broad path length
distribution in multiple scattering media
\cite{genack_relationship_1990}. These temporal and spatial
distortions are not separable \cite{lemoult_manipulating_2009}.
\newline\indent%
There is a strong interest in improving applications of ultrashort
laser pulses in complex scattering media. Phase conjugation has been
applied to spatially focus light from a short-pulse laser source
through a thin scattering layer \cite{hsieh_digital_2010}.
Similarly, phase conjugation is applied to correct distortions of
the ballistic wave front to improve the resolution of two photon
microscopy \cite{rueckel_adaptive_2006}. Coherent control of
two-photon excitation through scattering biological tissue has been
demonstrated \cite{dela_cruz_use_2004}. Those experiments share the
common limitation that the control is limited only to those photons
that take the shortest paths through the disordered media and arrive
a the target volume without being multiply scattered.
\newline\indent
Recently it was demonstrated that random scattering can actually be
beneficial rather than detrimental for the performance of optical
systems. Applying a shaped wave front of monochromatic light to a
strongly scattering medium, Vellekoop et al. achieved spatial
control over the scattered light \cite{vellekoop_focusing_2007}. In
fact, the insertion of an opaque sample after a lens has allowed
focusing beyond the diffraction limit of the lens
\cite{vellekoopi._m._exploiting_2010}. These findings have opened
new possibilities for imaging in optically thick biological matter
\cite{mcdowell_turbidity_2010} and allow trapping particles through
turbid media \cite{cizmar_in_2010}. Related techniques which allow
coherent focusing in scattering media are known from ultrasound
\cite{derode_robust_1995} and microwaves
\cite{lerosey_focusing_2007}. The frequency of those types of waves
is low enough that electronic transducers can be used to time
reverse waves, which redirects the waves towards their source. This
technique has successfully helped to improve imaging resolution
\cite{fink_multiwave_2010} and communication bandwidth
\cite{simon_communication_2001}.
\newline\indent%
In this Letter we report the first experimental demonstration of
combined spatial and temporal control of light trajectories through
opaque media via spatial wavefront shaping. We apply our method to
create an ultrashort pulse from the light transmitted through a
strongly scattering medium. We can control the amount of time the
optimized pulse stays in the sample and thereby select the path
length of the light through the medium. The efficiency of our method
is independent of the time delay, allowing for ultrashort pulses to
be formed even from very long paths in the sample.
%
\newline\indent
%
\begin{figure}[ht]
  \includegraphics[width=0.48\textwidth]{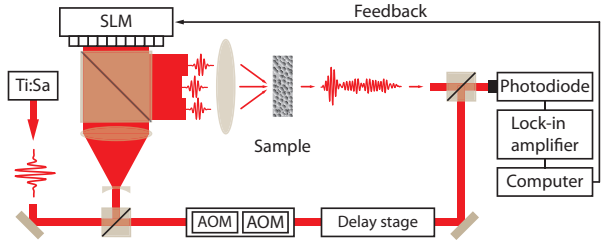}\\
  \caption{Experimental setup. A beam from a mode-locked Titanium:Sapphire (Ti:Sa) laser is coupled into a heterodyne Mach-Zehnder interferometer.
  In the signal branch the beam reflects off a spatial light modulator (SLM) and is subsequently focused onto the sample. The light in
  the reference branch is frequency shifted by 40 kHz by two acousto-optical modulators (AOM) and then passes through a motorized delay
  line. The signal from the detector is filtered by a lock-in
amplifier (LIA) and recorded by the Computer.}\label{FigSetup}
\end{figure}
The experimental realization can be summarized as follows (Figure
\ref{FigSetup}). An short-pulse light source illuminates a
phase-only spatial light modulator (SLM), which can alter the phase
of the light reflected from its surface. The SLM pixels are grouped
into N independent segments each of which induces a controllable
phase shift $\Delta\Phi_i$, which can be considered constant over
the bandwidth of the laser. The scattering sample is placed in the
Fourier plane of the SLM.
Both SLM and sample are embedded in the signal arm of a heterodyne
Mach-Zehnder-type interferometer \cite{sandtke_novel_2008}. The
heterodyne signal exactly corresponds to the cross-correlation of
the of the forward scattered pulse with the reference pulse, which
is delayed by a variable time delay $\tau$. The reference pulse is
close to Fourier-limited, so that amplitude and phase of the
transmitted pulse will not change significantly over the duration of
the reference pulse. The heterodyne signal is then effectively an
instantaneous measurement of the transmitted electric field.
This signal serves as a feedback for an optimization algorithm,
which programs the SLM. The time delay of the reference pulse is
adjusted by an automated delay stage allowing optimization at a
desired point in time $\tau_{opt}$. An extended description of the
setup including technical details has been included in the
supplementary material.
\begin{figure}[ht!!!]
  \includegraphics[width=0.48\textwidth]{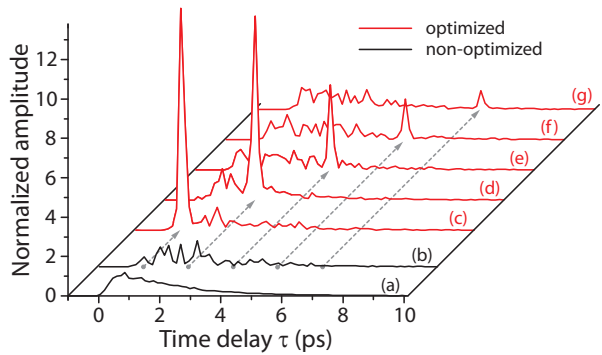}\\
  \caption{Optimized and random speckle pulses.
  {(a)} Amplitude of heterodyne signal of a non-optimized pulse as a function of time delay, averaged over 50 random speckle pulses.
  {(b)} Typical single random speckle pulse.
  {(c)-(g)} Amplitudes of single pulses after
  optimization at different time delays which are indicated by the dashed arrows. The optimization has been
  performed by dividing the SLM into 300 segments. The optimization generates strong, short pulses from diffuse light. The zero delay position is at the
maximum amplitude with no sample. The plotted curves have been
normalized to the maximum of the average non-optimized heterodyne
signal (factor 1.53 mV$^{-1}$).}
  \label{FigSingleSpeckle}
\end{figure}
\newline\indent%
The principle of the experiment can be described as follows. Light
reflected from a single segment on the SLM is transmitted through
the sample, giving rise to the field $E_i(t)$ at the detector. Its
phase can be modified by a time-independent phase shift
$\Delta\Phi_i$ via the SLM.
 The total field scattered into the detector
$E_{out}(t)$ is therefore given by the sum over all segments
\begin{equation}\label{Eq_LinCom}
    E_{{\rm out}}(t) = \sum\limits_{i=1}^N E_{i}(t)
    e^{i\Delta\Phi_i}.
\end{equation}
\newline\indent%
Multiple scattering allows us to assume that the contributions
$E_{i}(t)$ from the different segments at every single point in time
$t$ are uncorrelated random variables with Rayleigh distributed
amplitudes $|E_{i}(t)|$ and uniformly distributed phases
$\Phi_{i}(t)$ \cite{van_tiggelen_delay-time_1999}. For the
non-optimized case, the resulting field $E_{\rm out}(t)$ can be
viewed as the result of a random walk in the complex field plane.
After the optimization, all contributions are in phase, adding up
constructively. The average amplitude enhancement is given by
\cite{vellekoop_focusing_2007}
\begin{eqnarray}
  \langle\alpha\rangle &=& \frac{\langle |E_{\rm opt}|\rangle_{\rm rms}}{\langle |E_{\rm rnd}|\rangle_{\rm rms}}\nonumber\\
   &=& (\frac{\pi}{4}(N-1)+1)^{1/2} \approx (\frac{\pi}{4}N)^{1/2}.
   \label{Eq_Enhancement}
\end{eqnarray}
The average intensity enhancement $\eta$  can be obtained by $\eta =
\alpha^2$.
\newline\indent%
The non-optimized data was obtained by setting random phase values
to the SLM segments. The optimization algorithm adjusts the phase
shifts $\Delta\Phi_i$ such that the amplitude of the heterodyne
signal is maximized. We performed the optimization at 20 equidistant
time delays $\tau_{\rm opt}$ between \mbox{-1.05~ps} to +13.6~ps.
For each $\tau_{\rm opt}$, the optimization was performed four
times, with N = 12, 48, 192 and 300 segments, respectively, each
time starting from a new random phase pattern.
\newline\indent%
Our main result is displayed in Figure \ref{FigSingleSpeckle},
showing the amplitudes of both the non-optimized (black lines) and
the optimized (red lines) pulses for different time delays
$\tau_{\rm opt}$ and N = 300 segments on the SLM.  The optimized
amplitudes show sharp, distinct peaks with dramatically increased
amplitudes at the desired time delay. We can control the amount of
time the optimized pulses stay in the sample by the time delay
$\tau_{\rm opt}$, and by that we control the path length of the
pulses through the sample. Note that the heterodyne signal is
proportional to the field amplitude, the intensities exhibit even
more pronounced optimized peaks.
\begin{figure}[ht]
  \includegraphics[width=0.34\textwidth]{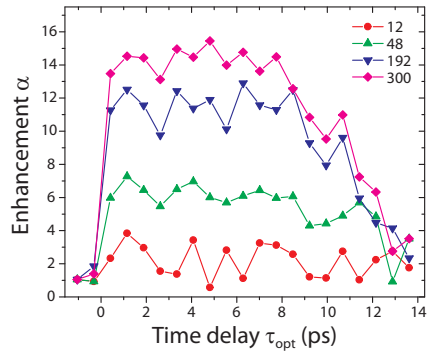}
  \caption{Enhancement $\alpha$ versus selected time delay $\tau_{\rm opt}$ for different number of segments N on the spatial light modulator.}
  \label{FigOptimizationSummary}
\end{figure}
\newline\indent%
The enhancement $\alpha$ versus time delay $\tau_{\rm opt}$ is shown
in Figure \ref{FigOptimizationSummary}. Its magnitude, depending on
the number of segments on the SLM, is constant from zero to several
picoseconds time delay. This result shows that our method works for
short light paths as well as for light paths more than ten times
longer than the sample thickness.
\newline\indent%
For long time delays a continuous decrease of $\alpha$ is observed,
which is related to the noise level of the experiment. We include a
quantitative analysis of this effect in the supplementary material.
\begin{figure}[ht!!!]
  \includegraphics[width=0.34\textwidth]{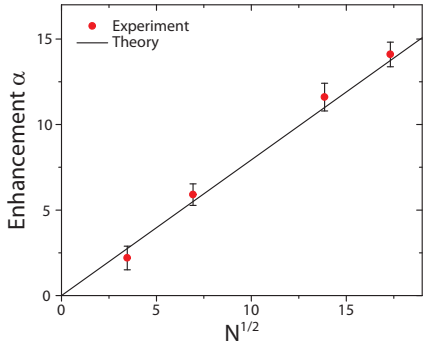}\\
  \caption{Average amplitude enhancement $\alpha$ (red dots) as a function of the square root of the number of segments $N$. The solid line is \
  given by the expected $\alpha
=\,0.90\,(\frac{\pi}{4}N)^{1/2}$, without any free parameter.}
  \label{FigPixelVsEnhancement}
\end{figure}
\newline\indent%
Figure \ref{FigPixelVsEnhancement} shows the average enhancement in
the constant regime in Figure \ref{FigOptimizationSummary} versus
$N$ together with the enhancement expected from theory
(Eq.\ref{Eq_Enhancement}), $\langle\alpha\rangle
=\,\sigma(\frac{\pi}{4}N)^{1/2}$. The prefactor $\sigma\,=\,0.90$
corrects for the non-uniform illumination of the SLM surface with a
truncated Gaussian beam, which effectively leads to a reduction of
the number of used segments (see supplementary material). Our model
matches the data well with no adjustable parameters.
\newline\indent%
We have investigated the duration of optimized pulses in detail.
Figure \ref{FigOptimizedPeakWithFit} shows the heterodyne signal of
three typical optimized pulses in a detailed scan of the time delay
around the respectively chosen optimization time $\tau_{opt}$. The
average width of the optimized peaks (full width at half maximum)
$\Delta\tau_{\rm opt}={\rm 190\pm 7\,fs}$, shows no dependence on
time delay range between zero and 10~ps. The heterodyne signal is
given by the amplitude of the optimized pulses, convoluted with the
bandwidth-limited reference pulses. By deconvolution we calculated
the field amplitude of the optimized pulses and determined their
(intensity) pulse duration $\Delta\rm t_{opt}$. The optimized pulses
have a duration $\Delta\rm t_{opt}$~=~115~fs. For comparison, the
input pulses have a transform-limited duration of $\Delta\rm
t_{in}$~=~64~fs. This raises the question whether the lengthening of
the transmitted pulse is caused by spectral narrowing or by
remaining fluctuations of the spectral phase. Numerical simulations,
which are appended in the supplementary material, show that the
answer is dependent on the number of segments N. For a low number of
segments, the spectrum of the optimized pulse is random with a very
short frequency correlation, but overall retains the shape and the
width of the Gaussian input spectrum. The optimized spectral phase
is not flat so that the pulses do not reach the bandwidth limit. For
an increasing number of segments, the duration of the optimized
pulse converges to its Fourier-limit. The spectrum has a smooth
Gaussian shape, but a bandwidth narrower than the input spectrum.
The method is capable of creating bandwidth-limited pulses, but
since it is based on linear interferometry, the adaption to other
pulse shapes is equally
possible.%
\newline\indent%
The time-integrated intensity (energy) of the pulse with the highest
peak depicted in Figure \ref{FigSingleSpeckle} is 13.5 times higher
than the energy of the average non-optimized pulse. In addition to
the temporal optimization, overall more light is transmitted into
the detected channel, demonstrating that the scattered light is
controlled spatially and temporally. Our method exploits the mixing
of spatial and temporal degrees of freedom by the random medium
\cite{lemoult_manipulating_2009}, to control the transmitted light
in two spatial and one temporal dimension by only controlling
spatial degrees of freedom on the two-dimensional SLM. On the one
hand, the conversion of spatial degrees of freedom into temporal
ones comes to the price of a speckle background, which on the other
hand is easily outweighed by the enormous number of degrees of
freedom provided by state-of-the-art SLMs. The large number of
controllable spatial degrees of freedom is a great advantage over
frequency domain pulse shaping techniques. Spatiotemporal control of
the light field allows a far more generalized application of present
coherent control schemes and marks a further step towards optical
time reversal.%
\newline\indent%
In the experimental realization presented here, we optimized the
pulse front using linear interferometry as feedback signal. The
optimization of a non-linear response, like second-harmonic
generation, will also lead to a comparably optimized pulse
\cite{yelin_adaptive_1997}. Using a nanoparticle with a non-linear
emission response \cite{hsieh_digital_2010} then enables the
focusing of ultrashort pulses inside complex scattering media. We
envision that our method can improve approaches for selective cell
destruction in tissue \cite{loo_immunotargeted_2005}. In view of its
potential for sharp focussing, it has potential for nanofabrication,
nanosurgery and other micromanipulation techniques.
\newline\indent%
Up to now we have not discussed the spatial extent of the optimized
pulse. We use an aperture to select a single speckle spot in the
Fourier plane of the sample for optimization, which corresponds to
light transmitted into the forward direction. We know that
transmitted fields in adjacent speckle spots are uncorrelated
\cite{sebbah_waves_2001}, from which we can conclude that the
optimization is indeed limited to the selected area. An important
future direction would be to investigate the spatial extent of the
optimized pulse as a function of delay time. A combination with
spatial scanning allows the measurement of the transmission matrix
of the medium \cite{popoff_measuring_2006} in one temporal and two
spatial dimensions. For Anderson-localizing samples
\cite{anderson_absence_1958}, the size of the optimized speckle
should be strongly time-dependent \cite{skipetrov_dynamics_2006}.
\newline\indent%
In conclusion, we have experimentally demonstrated that spatial wave
front shaping of a pulse front incident on a strongly scattering
sample gives spatial and temporal control over the scattered light.
Our approach provides a new tool for fundamental studies of light
propagation and has potential for applications in sensing, nano- and
biophotonics.
\newline\indent%
We thank Timmo van der Beek for the sample fabrication, Kobus
Kuipers for providing the AOMs and Huib Bakker for helpful comments
on the manuscript. This work is part of the Industrial Partnership
Programme (IPP) Innovatie Physics for Oil and Gas (iPOG) of the
Stichting voor Fundamenteel Onderzoek der Materie (FOM), which is
supported financially by Nederlandse Organisatie voor
Wetenschappelijk Onderzoek (NWO). The IPP MFCL is co-financed by
Stichting Shell Research.

\end{document}